\documentclass[sigconf, 10pt]{acmart}

\usepackage{booktabs} 
\usepackage[T1]{fontenc}
\usepackage[utf8]{inputenc}
\usepackage[greek,english]{babel}
\usepackage{alphabeta}

\usepackage{algorithm}
\usepackage{algcompatible}

\algnewcommand{\LINECOMMENT}[1]{\STATE\(\triangleright\) #1}

\usepackage{caption}
\usepackage{subcaption}

\usepackage{xcolor}
\newcommand{\red}[1]{\textcolor{red}{#1}}
\newcommand{\blue}[1]{\textcolor{blue}{#1}}

\usepackage{paralist}

\setcopyright{rightsretained}

\usepackage{caption}
\usepackage{subcaption}
\usepackage{multirow}
\usepackage{tabularx}
\newcommand{\dimhatzo}[1]{{\color{gray} #1}}
\newcommand{\orange}[1]{\textcolor{orange}{(Dimitris: #1)}}

\acmConference[ ]{ }{ }{ }

\begin{document}
\title{IPLS : A Framework for Decentralized Federated~Learning}

\author{Christodoulos Pappas}
\affiliation{%
  \institution{University Of Thessaly}
}
\email{chrpappas@uth.gr}

\author{Dimitris Chatzopoulos}
\affiliation{%
  \institution{HKUST}
  \streetaddress{}
  \city{} 
  \state{} 
}
\email{dcab@cse.ust.hk}

\author{Spyros Lalis}
\affiliation{%
  \institution{University of Thessaly}
  \streetaddress{}
  \city{} 
  \state{} 
}
\email{lalis@uth.gr}

\author{Manolis Vavalis}
\affiliation{%
  \institution{University of Thessaly}
  \streetaddress{}
  \city{} 
  \state{} 
}
\email{mav@uth.gr}
\begin{abstract}
    The proliferation of resourceful mobile devices that store rich, multidimensional and privacy-sensitive user data motivate the design of federated learning (FL), a machine-learning (ML) paradigm that enables mobile devices to produce an ML model without sharing their data. However, the majority of the existing FL frameworks rely on centralized entities. In this work, we introduce IPLS, a fully decentralized federated learning framework that is partially based on the interplanetary file system (IPFS). By using IPLS and connecting into the corresponding private IPFS network, any party can initiate the training process of an ML model or join an ongoing training process that has already been started by another party. IPLS scales with the number of participants, is robust against intermittent connectivity and dynamic participant departures/arrivals, requires minimal resources, and guarantees that the accuracy of the trained model quickly converges to that of a centralized FL framework with an accuracy drop of less than 1\textperthousand.
\end{abstract}

\maketitle
\section{Introduction}

Federated learning (FL) is a recently proposed ML paradigm that allows entities which store locally, potentially privacy-sensitive, data to train models collectively~\cite{10.1145/3298981}. The most prominent example is Google Keyboard that uses metadata from users' typing to propose next words or to auto-correct typed words, while preserving users privacy~\cite{DBLP:journals/corr/abs-1812-02903}. 

In traditional FL, a \textit{centralised server} orchestrates the training process by 
\begin{inparaenum}[(i)]
\item determining the type of the model (e.g., a deep neural network) to be trained by several agents using the same loss function and optimisation algorithm (e.g., stochastic gradient descent~\cite{robbins1951stochastic}), 
\item registering the interested agents and recording their contact information in order to be able to communicate with them directly, 
\item randomly sampling a subset of the agents for the next training round, 
\item sending to each of these agents the most updated values of the global model parameters, and 
\item aggregating the individual agent contributions in order to update the global model parameters to be used in the next training round.
\end{inparaenum}

\begin{figure*}[t]
    \centering
    \begin{subfigure}{\columnwidth}
        \centering
        \includegraphics[width=\columnwidth]{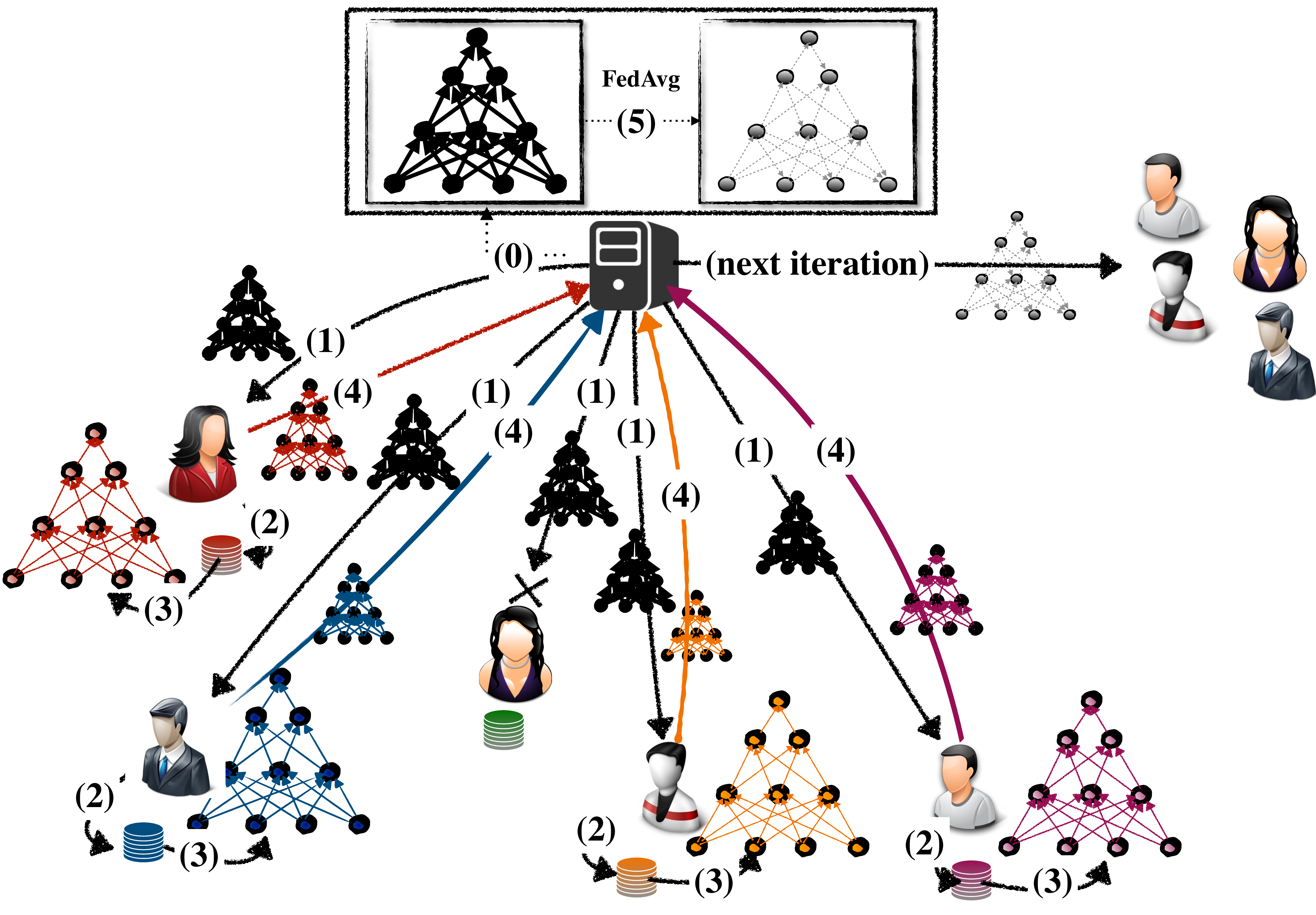}
        \caption{Centralised FL.}
        \label{fig:trad_FL}
    \end{subfigure}
    \begin{subfigure}{\columnwidth}
        \centering
        \includegraphics[width=\columnwidth]{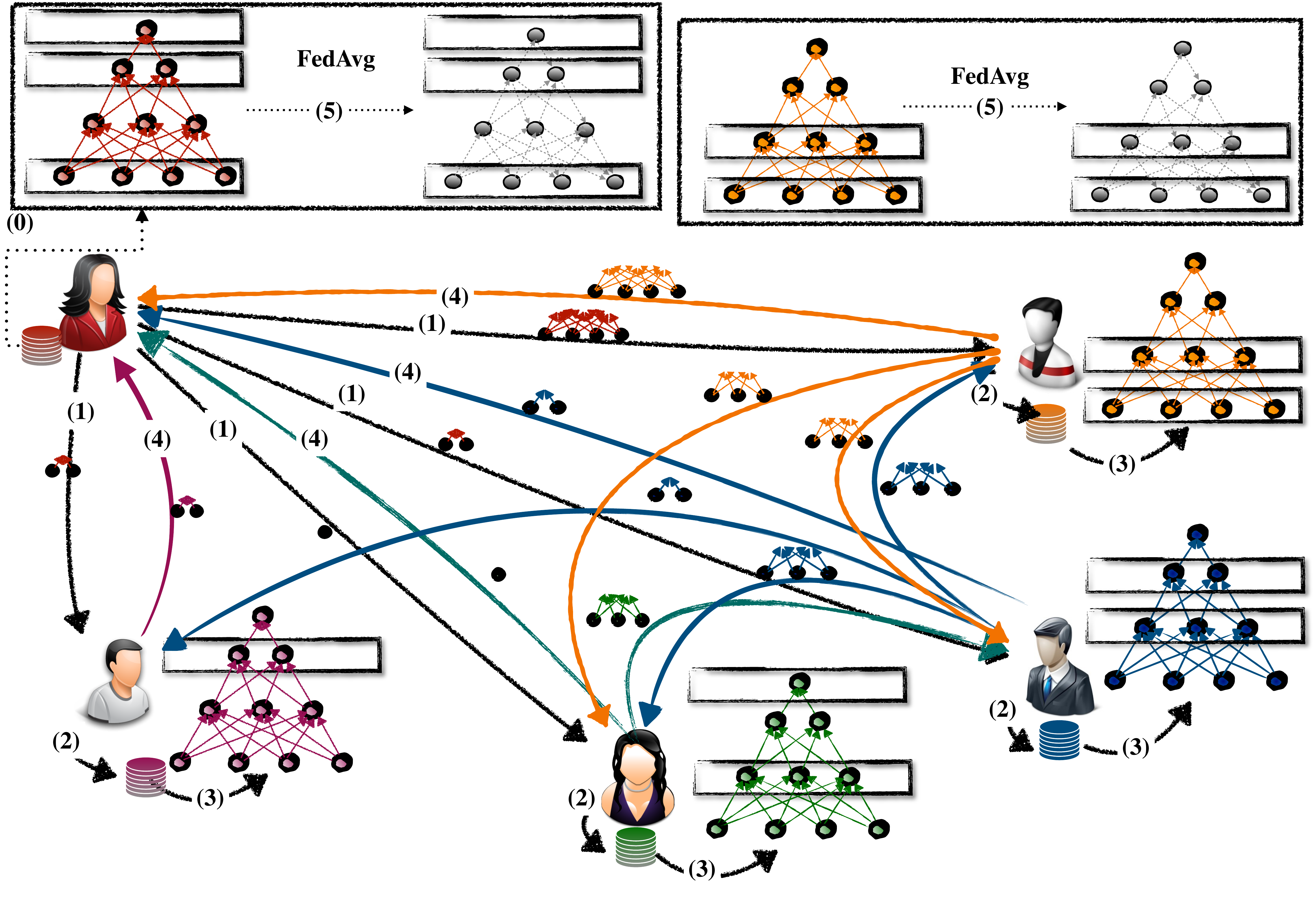}
    \caption{Decentralised FL.}
    \label{fig:dec_FL}
    \end{subfigure}
    \caption{In centralized FL (Figure (a)), each agent sends the updated model to the server, the server produces the new model, and begins a new training phase. In IPLS (Figure (b)) each agent is responsible for some partitions of the model and agents interact with each other by exchanging partitioned gradients or model updates.}
    \label{fig:tradvsdec}
\end{figure*}

Model, loss function and algorithm determination, as well as the registration of the agents, are components of an initialisation process, which takes place before the training process. The training process, depicted in Figure~\ref{fig:trad_FL}, takes place in rounds, until the global parameters converge. 
In each round, the chosen agents  receive the global parameters from the server, execute the optimisation algorithm for a predetermined period (specified in time units of number of iterations) 
using only locally stored data. When the period expires, each agent calculates the difference between the locally trained model and the global model that was received from the server, and reports this difference back to the server.

In decentralised FL, illustrated in Figure~\ref{fig:dec_FL}, the agents \emph{collectively} train a model in a peer-to-peer fashion \emph{without} the assistance of a server. Any agent can initiate the training process by specifying the model, the loss function and the employed algorithm. Then, interested agents may register and participate in the training process. In contrast to the centralised setting, where only the server is responsible for storing, updating and broadcasting the model to the participating agents, in decentralised FL, the model is split in multiple partitions 
that are replicated on multiple agents. 
For example, a model using a neural network of 100 layers~\cite{he2016deep} can be split in 10 partitions of 10 layers each. As a consequence, each agent is responsible for storing a part of the model, updating the corresponding parameters and communicating them to the agents working on the other partitions. Notably, all agents that are responsible for the same partition need to agree on 
the same values, by running a suitable 
aggregation protocol~\cite{10.1145/3133956.3133982}. In this work we consider asynchronous aggregation protocols that do not guarantee that an agreement has to be reached after each round.

It is easy to see that traditional FL 
has a single point of failure and any 
unavailability of the central server will cause an immediate and complete disruption of the training process. Also, the server needs to have reliable and high-bandwidth communication links with the agents in order to support the transfer of potentially voluminous data with all of them. Last but not least, 
the server needs to be trusted by all agents.
For example, in a scenario where multiple users of mobile devices want to train collectively a model that recognises emotions through speech~\cite{9111050}, they need to hire a server with quality of service guarantees to orchestrate the process. An attractive alternative is to use the decentralised approach to train the model while relying only on their own resources. 




Inspired by the design and functionalities of the Interplanetary File System (IPFS)~\cite{benet2014ipfs}, this paper introduces an decentralized FL framework, named Interplanetary Learning System (IPLS), which allows a large number of potentially mobile agents to collaborate in the training of a model without relying on any central entity. The main contributions are:
\begin{inparaenum}
\item We propose a new algorithm for decentralized FL based on shared memory, which has very similar convergence rate and network traffic with centralized FL. 
\item We present a concrete implementation, in the form of a middleware atop IPFS, which can be used through a structured API by anyone who wishes to train an ML model, without having to hire and maintain a centralized service (as done in traditional ML systems). 
\item We evaluate the effectiveness of IPLS via a set of experiments, showing that it can scale to a large number of nodes with satisfactory accuracy and convergence compared to a centralized approach. 
\end{inparaenum}

The rest of the paper is structured as follows: In Section~\ref{sec:IPLS}, we introduce IPLS in detail; in Section~\ref{sec:evaluation} we evaluate the performance of IPLS; in Section~\ref{sec:relatedWork} we compare IPLS to related work and, finally, Section~\ref{sec:conclusion} concludes the paper and points to future research directions.

\section{Interplanetary Learning System}\label{sec:IPLS}
The design of IPLS is based on two assumptions in order to guarantee four desirable properties.

\noindent\textbf{Assumptions.} We assume that every agent that participates in the training of a model using~IPLS:

\textbf{1) Mobile.} Agents are mobile (e.g., autonomous vehicles or smartphones) and in full control of their own mobility.

\textbf{2) Availability.} Agents may get disconnected from the Internet and their peers or may terminate an IPLS-based training process to save energy or other resources. We furthermore assume that nodes remain unavailable only for a short while, unless they exhibit a permanent failure or leave the training process.


\noindent\textbf{Properties.} We design IPLS in such a way to guarantee the following properties:

    \textbf{1) Model training convergence.} The global parameters converge to a set of values and the accuracy of the model is very close to that of a model that is trained in a centralised fashion with the same data.
    
    \textbf{2) Scalability.} The produced traffic by IPLS increases sub-linearly to an increase in the participating agents. Moreover the increasing participation does not affect the communication complexity of an agent. 
    
    \textbf{3) Fault-tolerance.} Even if a fraction of the agents leaves the process unexpectedly, the training process terminates successfully, the global parameters converge to a set of values and the accuracy of the model is very close to that of a model that is trained in a centralised fashion with the same data.
    
    \textbf{4) Lightweight storage requirements.}  Besides, the locally stored data each agent owns and uses during training, IPLS requires relatively little space to store part of the model.


\subsection{Training a model with IPLS}

Given a model $M$, with weight parameters $W$, and a set of agents $\mathcal{A}$ with each agent $a_i \in \mathcal{A}$ owning a private dataset $d_i$, we next describe how IPLS trains $M$ in a decentralised way. Every IPLS agent runs an IPFS daemon and utilizes IPFS to exchange data with other agents.

\smallskip\noindent\textbf{Initialisation phase.}
Any agent can initiate the training process by determining \textit{(i)} the characteristics of $M$, i.e, the topology of the model 
(e.g., ResNet~\cite{he2016deep}),
\textit{(ii)} an optimisation algorithm, and \textit{(iii)} a loss function $L$, which will be used to optimise the weights of $M$. IPLS uses the pub/sub module~\cite{10.1145/41457.37515} of IPFS to notify agents about the initialisation of a training process and invite them to express their interest.

\smallskip\noindent\textbf{Model partitioning and distribution.} Depending on the size of $M$, $W$ can be split into multiple partitions.
Each agent can be assigned multiple partitions. The partitions need to be both distributed and replicated among the agents so that any agent can find,for every partition she does not store locally, at least one agent that is responsible for it ,with high probability.
Formally, for $K$ partitions we have $W = \bigcup_{k = 1}^{K} w_{k}$ while agent $a_i$ stores partitions \textbf{$k_i$}. IPLS implements a mechanism for the distribution of the partitions that is based on the storage space, $s_i$, each agent $a_i$ shares and on two tuning parameters $\pi$ and $\rho$.
$\pi$ denotes the minimum number of partitions an agent can store and $\rho$ the maximum number of times a partition can be replicated.
At the beginning, the agent that initiated the training process stores all the partitions. Whenever another agent expresses her interest to participate, she gets $\pi$ partitions from the agent she has access to and stores most of the partitions. If multiple agents have the same number of partitions, the agent selects the $\pi$ least replicated partitions. 

\smallskip\noindent\textbf{Partitioning example.} Agent 1 initiates the process and stores all 6 partitions, $k_1 = 1,2,3,4,5,6$ while $\pi = 4$ and $\rho=2$. Agent 2 expresses her interest to participate and stores partitions $k_2 = 3, 4, 5, 6$ while agent 1 remains responsible for partitions $k_1 = 1,2,3,4$. Next, agent 3 expresses her interest to participate and stores partitions $k_3 = 1,2,5,6$. Any other agent that wishes to participate cannot replicate any partition since all of them have been replicated twice and $\rho=2$. New agents cannot store any partition because they will violate the restrictions $\pi = 4$ and $\rho=2$ put. 

Ideally, all the partitions will be replicated $\rho$ times. 
Model partitioning and distribution are parts of the initialisation phase. By the end of it, each agent knows sufficient IPFS addresses to retrieve all the partitions and the addresses of the agents who are store the same partitions as her. 

\smallskip\noindent\textbf{Training phase.} During the training phase each agent initially contacts enough agents to collect the global parameters. The number of the contacts depends on the number of the partitions she stores locally and the partitions she needs in order to get the whole model. Next, each agent, $a_i \in \mathcal{A}$ uses her locally stored data, $d_i$, the predetermined optimisation algorithm and the loss function to update the model parameters by running the algorithm for a given number of iterations. Finally, each agent calculates the difference between the updated parameters and the ones she retrieved before starting the optimisation and informs the agents from which she retrieved each partition. For every partition, all agents who are responsible to store it exchange the newly calculated values for the parameters together with the identifiers of the agents that submitted them in order to calculate the new global parameters.  

\smallskip\noindent\textbf{Communication complexity example.} Assuming $\rho = 1$ and partitions of equal size (i.e., $w_1=w_2=\ldots=w_K=w$), each agent $a_i$ has to send an update to agent $a_j$, $i \neq j$, of size $k_j w$. Thus, the updates send by agent $a_i$ are $\sum_{a_j \in \mathcal{A},j \neq i}{k_j w} = (K-k_i)w < W$, which are equal to the received updates. Thus the data communicated on each round round are less than $ |\mathcal{A}|(updates\_send + updates\_recv) \leq 2\mathcal{A}|M|$ which have the same volume as in traditional FL.

\smallskip\noindent\textbf{The impact of $\pi$ and $\rho$.} 
The difficulty for an agent to retrieve a partition that is not stored locally increases when $\pi$ and $\rho$ are small since fewer agents can provide the partitions. On the other hand, higher values of $\pi$ and $\rho$ increase the number of messages the agents need to exchange in order to update the global parameters. For example, if $\pi = 1$ and $\rho = 1$, only $K$ agents will store a partition and only one agent will be responsible for each partition. In this case, every agent needs to communicate with each of the the $K$ agents to get the global parameters and inform them about the produced updates by the end of the training round. For higher values of $\rho$, the agents that store the same partition need to reach a consensus in order to produce the new global parameters because each of them only retrieves the updates from the agents to which she has send the global parameters. However, higher values of $\rho$ increase the robustness of IPLS because whenever an agent is not available, the other agents have alternatives. 
The higher the value of $\rho$, the more decentralised IPLS is since its operation is less dependent on specific agents. However, for small values of $|\mathcal{A}|$ the communication overhead for updating the global parameters increases. In reality $\rho$ naturally increases as the participation increases and vice versa.


\subsection{IPLS API}
IPLS is build atop IPFS, a fully decentralized peer-to-peer file system with a pub/sub functionality that assists agents on communicating with each other. IPLS offers an API of four methods to anyone who wants to participate in the training of ML models in a decentralised way: \texttt{Init}, \texttt{UpdateModel}, \texttt{LoadModel}, and \texttt{Terminate}. Algorithm~\ref{alg:updateLocalModel} shows how the first three are used during the training of the model while the fourth one is used by agents who wants to quit training.

\smallskip\noindent\texttt{Init(String \emph{IPFS\_Path}, List \emph{Bootstrapers}):} implements the initialisation phase. It first initializes the IPFS daemon, using its IPFS\_Path. After that it broadcasts, using the pub/sub, the required communication addresses, a description of the characteristics of $M$, $L$, $\pi$, $\rho$ and the optimisation algorithm the participating agents need to use. After that, \texttt{Init()} waits for responses from interested agents. These responses, contain the communication addresses of the agents and the partitions that they are responsible for and the storage they are willing to allocate for the training. After receiving those data from enough agents, she selects the partitions she will store locally by 
selecting partitions from agents who have more partitions than $\pi$ or the least replicated partitions. Finally she broadcasts the partition distribution to all the other agents. 
All the addresses are stored in a lookup table. 
Next it starts a middleware manager daemon, who is responsible for keeping $W$ up to date and deal with the mobility of the agents.

\floatname{algorithm}{Algorithm}

\begin{algorithm}
\caption{Runs on agent $a_i \in \mathcal{A}$}
\begin{algorithmic}[1]
\REQUIRE{IPFS Path, List of Bootstrapers}
\STATE create Map<PartitionID,Address> agents\_k
\STATE IPFS.init\_deamon(String IPFS\_Path)
\STATE IPFS.broadcast($M,L,\pi,\rho$,\textit{``SGD''})
\STATE agents\_k $\leftarrow$ \texttt{IPFS.receive(timeout)}
\STATE $k_i$ $\leftarrow$ (agents\_k.key - partitions) 
\IF{$k_i.size$< $\pi$}
\STATE $k_i$ $\leftarrow$ $k_i$+(max(overloaded) $\wedge$ min(replicated))
\ENDIF
\WHILE{accuracy < Threshold}
    \STATE $M$ = \texttt{LoadModel()}
    \STATE $\Delta W$ = $M$.fit($d_i$,SGD)
    \STATE  \texttt{UpdateModel($\Delta W$)}
\ENDWHILE
\ENSURE{Updated local model.}
\end{algorithmic}
\label{alg:updateLocalModel}
\end{algorithm}

\smallskip\noindent\texttt{UpdateModel(Vector Gradient):} By the termination of the optimisation algorithm, agents call \texttt{UpdateModel()} to update $W$. Whenever this method is invoked, $W$ gets divided and organized into the corresponding pieces, and then for each partition performs a lookup in order to find agents that are responsible for a given partition. There can be many criteria for choosing the suitable agent, such as locality, connectivity, trust, load, power level etc. After selecting the appropriate agents, \texttt{UpdateModel()} sends the requests with each one containing the partition ID and the gradients sub-vector, and waits for the replies. The reply contains the partition ID and the updated sub-vector. Finally the received updated sub-vectors are stored in a cache for future use. Upon receiving an update for a partition $k$, $\delta_k$, an agent must update her it by subtracting $w_k$ with the $\delta_k$, multiple by a weight factor $\epsilon$. Assuming that she received an update from $r$ agents for $w_k$ in the last iteration, then she updates that weight factor by: $\epsilon \leftarrow \alpha\epsilon + (1-\alpha)\frac{1}{r}$ , where $\alpha \in (0,1)$.  

\smallskip\noindent\texttt{LoadModel():} This method combines and returns the cached global model received in by \texttt{UpdateModel()} method.

\smallskip\noindent\texttt{Terminate():} Whenever an agent calls this method, IPLS looks up for other agents based on their load and responsibilities and uploads to IPFS a file containing the model partitions for which she was responsible for and broadcasts a final message assigning to the selected agents her responsibilities. Upon receiving such message, the selected agents take the responsibility and also aggregate the downloaded weights of the corresponding partition to their own local weights to form a new global sub-vector.

\section{Performance Evaluation}\label{sec:evaluation}

\smallskip\noindent\textbf{Set up.} We implement a functional prototype of IPLS to measure its performance. For the simulation of the connectivity between the agents we use mininet\footnote{http://mininet.org/}. Each mininet node is an agent that uses IPLS in order to participate in the training of a model. Additionally, we set up a private IPFS network where every node runs as part of an IPLS agent. Agents communicate asynchronously and messages that are exchanged during a training iteration are probable to be lost or to be delivered after the start of the next training iteration. 

\smallskip\noindent\textbf{Dataset and model.} We use the MNIST dataset~\cite{deng2012mnist} that contains 60000 images that are categorised in 10 classes and trained a neural network with four layers ($785\times500\times100\times10$). We split MNIST into $|\mathcal{A}|$ parts, with uniformly distributed labels and assign to each agent a dataset of $60000/|\mathcal{A}|$ samples. Practically, when considering 10 agents, each agent has a dataset of $6000$ samples and the probability of one sample to belong to one class is the same for every agent.  


\begin{figure}[t]
    \centering
    \begin{subfigure}{0.7\columnwidth}
        \includegraphics[width=\columnwidth]{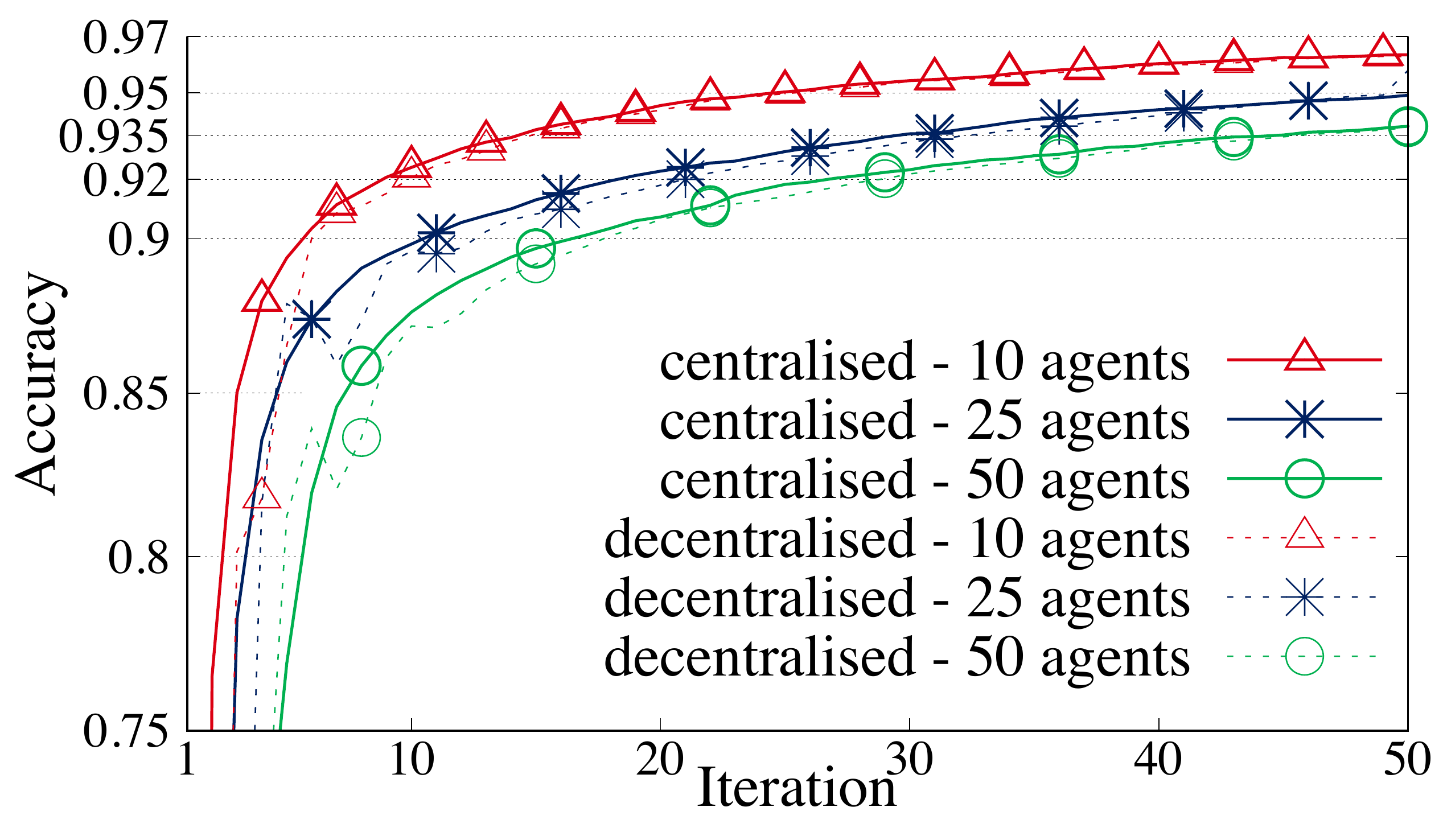}
        \caption{Centralised and decentralised agents}
        \label{fig:centvsdecentr}
    \end{subfigure}
    \begin{subfigure}{0.7\columnwidth}
        \includegraphics[width=\columnwidth]{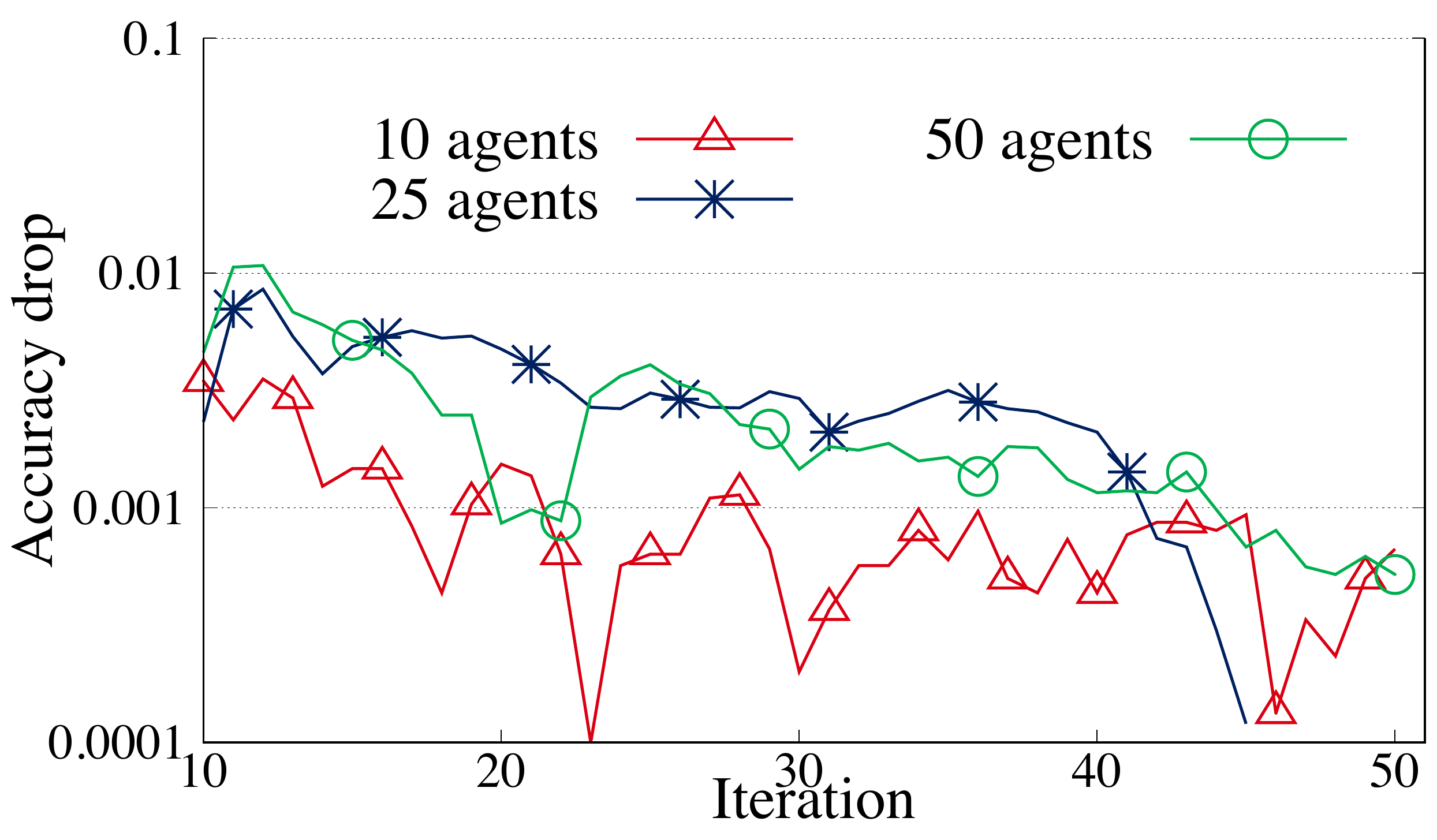}
        \caption{Accuracy ``loss'' due to decentralisation}
        \label{fig:accDrop}
    \end{subfigure}
    \caption{Model training convergence.}
    \label{fig:mtc}
\end{figure}

\begin{figure}[t]
    \centering
    \begin{subfigure}{0.7\columnwidth}
        \centering
        \includegraphics[width=\columnwidth]{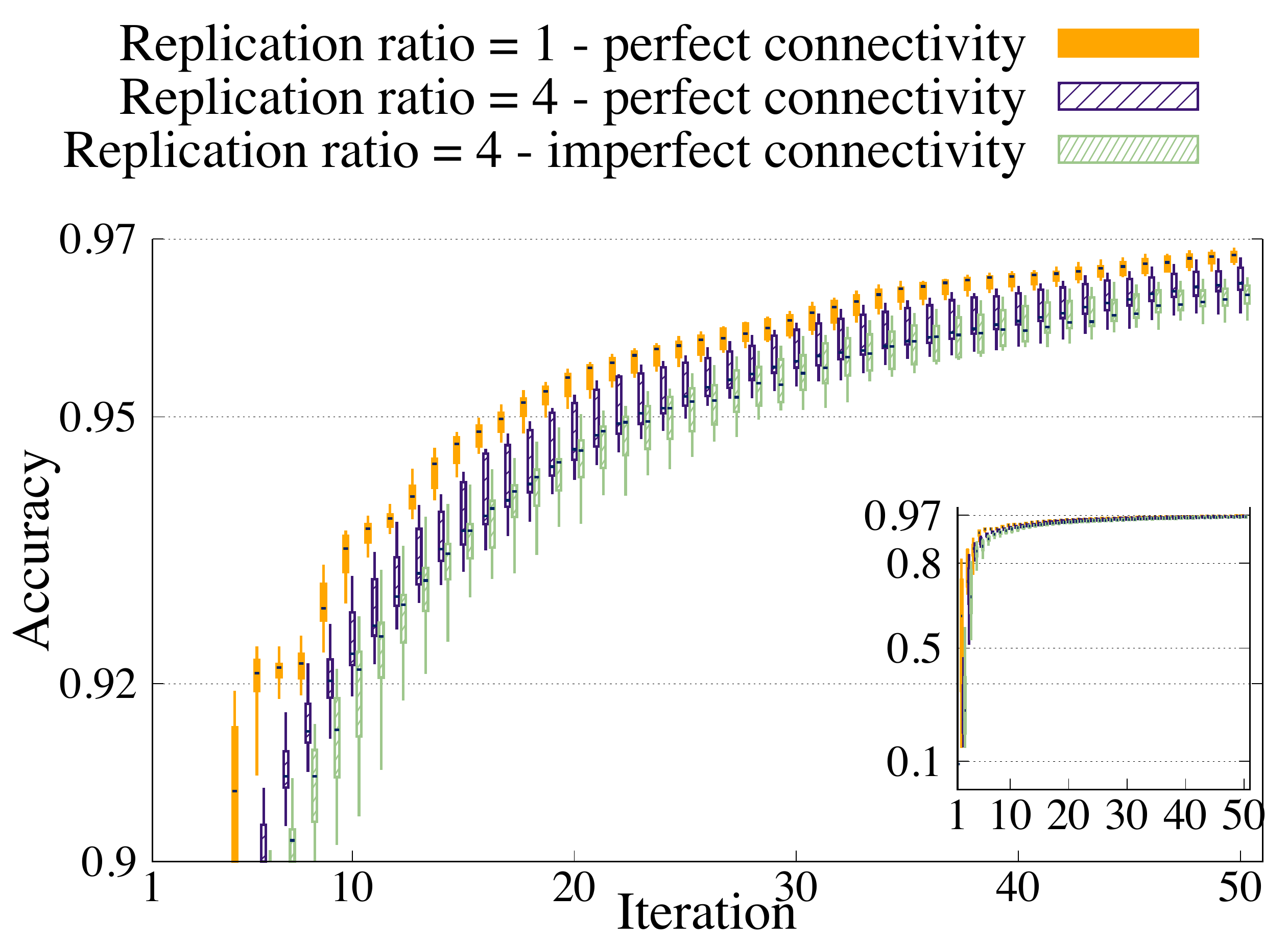}
        \caption{Replication ratio example}
        \label{fig:exp2}
    \end{subfigure}
    \begin{subfigure}{0.7\columnwidth}
        \centering
        \includegraphics[width=\columnwidth]{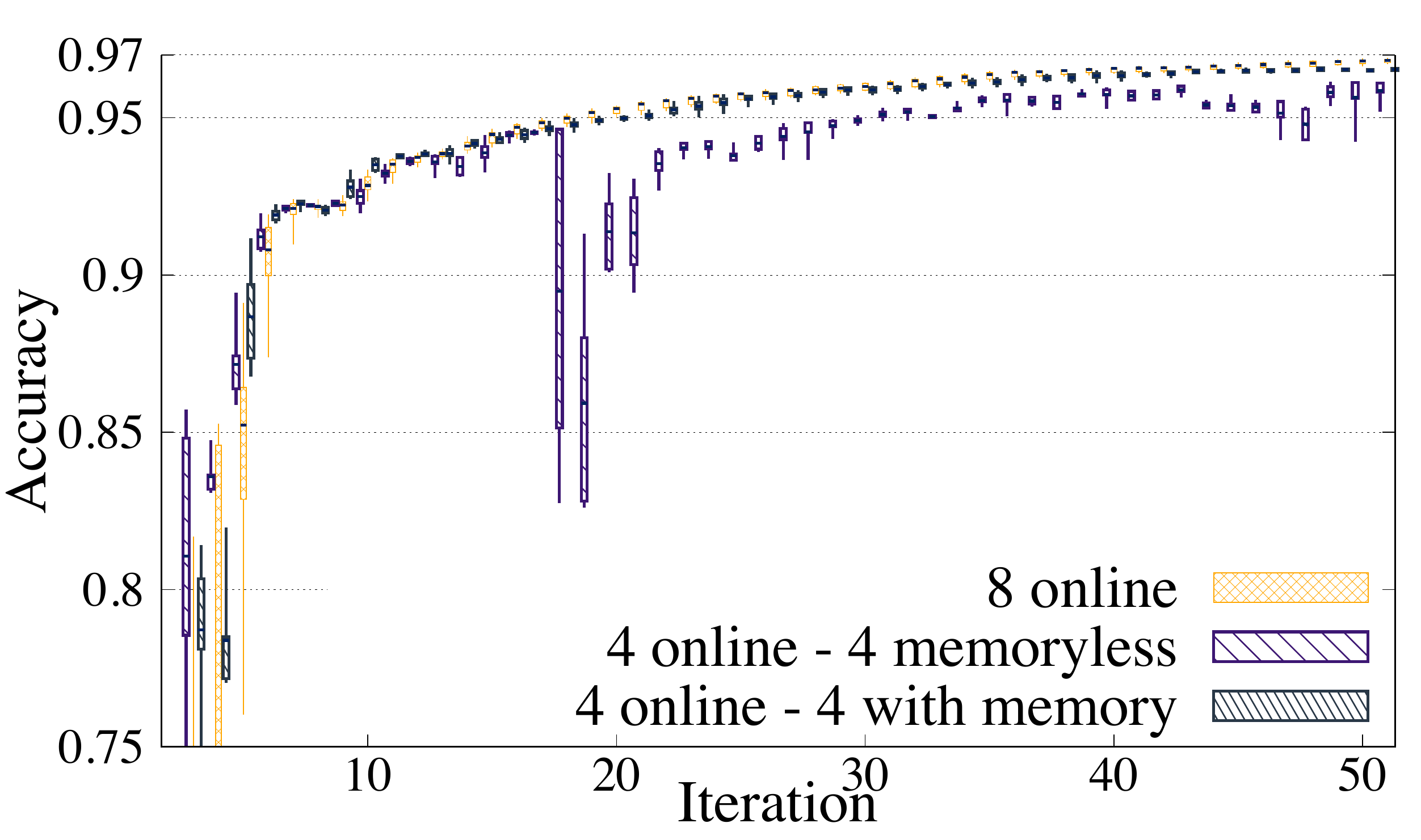}
        \caption{Disconnections.}
        \label{fig:exp3}
    \end{subfigure}
    \caption{Fault-tolerance.}
    \label{fig:ft}
\end{figure}

\smallskip\noindent\textbf{Experiments.} With focus on the justification of the four listed properties of IPLS, namely \textit{model training convergence}, \textit{scalability}, \textit{fault-tolerance} and \textit{lightweight storage requirements}, we design three experiments to present that \textit{(i)} the accuracy of a model trained with IPLS coverges to that of a model trained with centralised FL (Figure~\ref{fig:mtc}) and \textit{(ii)} IPLS tolerates agents' mobility and disconnections (Figure~\ref{fig:ft}). 

\textbf{Model training convergence.} First of all, we examine model training convergence by examining three scenarios with 10, 25 and 50 agents. Figure~\ref{fig:centvsdecentr} depicts the accuracy increase in all of them as the iterations increase as well as the convergence of IPLS to the centralised FL. Additionally, we confirm that if a fixed dataset is partitioned in fewer parts and given to less agents, the accuracy of the model is higher. This is explained by the fact that each agent has more data when updating her local model. Figure~\ref{fig:accDrop} shows the ``accuracy drop'' due to decentralisation, that after 40 iterations is less than 1\textperthousand.

\textbf{Fault-tolerance.} Next, we examine how the value of $\rho$ (i.e., replication ratio of partitions) impacts accuracy by considering three scenarios with $\rho = 1$ and perfect connectivity, $\rho = 4$ and perfect connectivity, and $\rho = 4$ and imperfect connectivity. Figure~\ref{fig:exp2} depicts the outcome of these three scenarios with 8 agents. First we note that the accuracy decreases when $\rho$ increases. This is justified by the fact that the agents who are responsible for the same partition do not synchronise in time to produce the correct global parameters. This is evident from the higher variance in the accuracy in Figure~\ref{fig:exp2}. This issue is treatable by increasing the time between two iterations and allowing time for synchronisation. Additionally, we see that the accuracy drops when the network conditions deteriorate. 


Last, we designed an experiment to examine the impact in the accuracy of the trained model whenever agents get disconnected for a while and then either start from the beginning (``memoryless training'') or continue from where they stopped (''training with memory''). We see that the accuracy of the model does not drop even when half of the agents have connectivity issues while in the case of agents with memory even the variation of the accuracy is not high. 


\textbf{Scalability and storage requirements.}
As described in Section~\ref{sec:IPLS}, the data sent and received by each agent is constant because on each communication round it sends and receives data of at most of the size of the model. With the replication of the partitions comes the issue of their synchronization and the aggregation of the replicated weights from each device holding the exact partition. IPLS uses the IPFS pub/sub for this aggregation. Every device holding that replication subscribes to its topic and listens for events. However the with pure pub/sub the larger the $\rho$ the more data an agent has to receive. IPLS has lightweight storage requirements as agents only need to store the models in which they participate in their training.




\section{Related Work}\label{sec:relatedWork} 




Existing decentralized FL systems are mostly based on gossiping schemes. For example, the authors of \cite{koloskova2019decentralized} and  \cite{DBLP:journals/corr/abs-1905-06731} implement the classic decentralized ML algorithm on which agents download the model from multiple neighbouring agents. An alternative approach is proposed by Ramanan~\textit{et al.}~\cite{DBLP:journals/corr/abs-1909-07452} who use a blockchain to aggregate agents' updates. However, their approach has several limitations related to the gas costs and the data size of each blockchain-based transaction.

Although the work of Hu~\textit{et al}~  \cite{DBLP:journals/corr/abs-1908-07782} is close to IPLS, since it also partitions the model into non overlapping segments, it differs heavily from IPLS because it is based on gossiping, and not on a distributed memory abstraction. Moreover, IPLS differs from \cite{koloskova2019decentralized,DBLP:journals/corr/abs-1905-06731,DBLP:journals/corr/abs-1909-07452} because it does not download the entire model from selected peers but only partitions of that model. The disadvantage of~\cite{koloskova2019decentralized,DBLP:journals/corr/abs-1905-06731,DBLP:journals/corr/abs-1909-07452} compared to IPLS, is that in order to gain better accuracy agents have to download the same partition from different agents. 
Compared to the aforementioned works, IPLS not only transmits significantly less data over the internet, but also reaches approximately the same convergence rate and accuracy as our centralized rival. Moreover given that IPLS is based on distributed shared memory, gives the API users more freedom to apply classic parallel optimization algorithms such as~\cite{recht2011hogwild} which can heavily reduce the communication complexity.

\section{Conclusion and Future work}\label{sec:conclusion}

The unavailability of a decentralized federated learning framework that can be used directly in mobile devices and especially smartphones motivated the development of IPLS. Although in an early stage, IPLS can be used to train models with the same convergence rate and the same traffic, as traditional FL frameworks. 

There are multiple directions towards which IPLS can be further developed. First of all, it needs to be  installed in different types of mobile devices in order to analyse extensively its energy needs and tested with as many as possible state-of-the-art models to examine its feasibility. A second improvement of IPLS is the integration of module that replaces pub/sub for simple read-only operations that require interactions between the agents. A fitting solution is the use of a smart contract that can be used as a directory service for all the model training activities that need more participants. Furthermore, a more sophisticated algorithm that allows agents to change the partitions for which they are responsible based on their bandwidth and their available resources can increase significantly the performance of IPLS because more updates will be delivered on time. Last but not least, IPLS should incorporate an incentive mechanism, similar to Filecoin~\cite{benet2017filecoin} and Flopcoin~\cite{chatzopoulos2017flopcoin}, to motivate mobile users to share their resources. 



\bibliographystyle{ACM-Reference-Format}
\bibliography{bib}


\begin{thebibliography}{16}


\ifx \showCODEN    \undefined \def \showCODEN     #1{\unskip}     \fi
\ifx \showDOI      \undefined \def \showDOI       #1{#1}\fi
\ifx \showISBNx    \undefined \def \showISBNx     #1{\unskip}     \fi
\ifx \showISBNxiii \undefined \def \showISBNxiii  #1{\unskip}     \fi
\ifx \showISSN     \undefined \def \showISSN      #1{\unskip}     \fi
\ifx \showLCCN     \undefined \def \showLCCN      #1{\unskip}     \fi
\ifx \shownote     \undefined \def \shownote      #1{#1}          \fi
\ifx \showarticletitle \undefined \def \showarticletitle #1{#1}   \fi
\ifx \showURL      \undefined \def \showURL       {\relax}        \fi
\providecommand\bibfield[2]{#2}
\providecommand\bibinfo[2]{#2}
\providecommand\natexlab[1]{#1}
\providecommand\showeprint[2][]{arXiv:#2}

\bibitem[\protect\citeauthoryear{Benet}{Benet}{2014}]%
        {benet2014ipfs}
\bibfield{author}{\bibinfo{person}{Juan Benet}.}
  \bibinfo{year}{2014}\natexlab{}.
\newblock \showarticletitle{Ipfs-content addressed, versioned, p2p file
  system}.
\newblock \bibinfo{journal}{\emph{arXiv preprint arXiv:1407.3561}}
  (\bibinfo{year}{2014}).
\newblock


\bibitem[\protect\citeauthoryear{Benet}{Benet}{2017}]%
        {benet2017filecoin}
\bibfield{author}{\bibinfo{person}{Juan Benet}.}
  \bibinfo{year}{2017}\natexlab{}.
\newblock \showarticletitle{Filecoin Research Roadmap for 2017}.
\newblock  (\bibinfo{year}{2017}).
\newblock


\bibitem[\protect\citeauthoryear{Birman and Joseph}{Birman and Joseph}{1987}]%
        {10.1145/41457.37515}
\bibfield{author}{\bibinfo{person}{K. Birman} {and} \bibinfo{person}{T.
  Joseph}.} \bibinfo{year}{1987}\natexlab{}.
\newblock \showarticletitle{Exploiting Virtual Synchrony in Distributed
  Systems}. In \bibinfo{booktitle}{\emph{Proceedings of the Eleventh ACM
  Symposium on Operating Systems Principles}} (Austin, Texas, USA)
  \emph{(\bibinfo{series}{SOSP '87})}. \bibinfo{pages}{123–138}.
\newblock
\showISBNx{089791242X}


\bibitem[\protect\citeauthoryear{Bonawitz, Ivanov, Kreuter, Marcedone, McMahan,
  Patel, Ramage, Segal, and Seth}{Bonawitz et~al\mbox{.}}{2017}]%
        {10.1145/3133956.3133982}
\bibfield{author}{\bibinfo{person}{Keith Bonawitz}, \bibinfo{person}{Vladimir
  Ivanov}, \bibinfo{person}{Ben Kreuter}, \bibinfo{person}{Antonio Marcedone},
  \bibinfo{person}{H.~Brendan McMahan}, \bibinfo{person}{Sarvar Patel},
  \bibinfo{person}{Daniel Ramage}, \bibinfo{person}{Aaron Segal}, {and}
  \bibinfo{person}{Karn Seth}.} \bibinfo{year}{2017}\natexlab{}.
\newblock \showarticletitle{Practical Secure Aggregation for Privacy-Preserving
  Machine Learning}. In \bibinfo{booktitle}{\emph{Proc. of the 2017 ACM SIGSAC
  Conference on Computer and Communications Security}}
  \emph{(\bibinfo{series}{CCS '17})}. \bibinfo{pages}{1175–1191}.
\newblock
\showISBNx{9781450349468}


\bibitem[\protect\citeauthoryear{Chatzopoulos, Ahmadi, Kosta, and
  Hui}{Chatzopoulos et~al\mbox{.}}{2017}]%
        {chatzopoulos2017flopcoin}
\bibfield{author}{\bibinfo{person}{Dimitris Chatzopoulos},
  \bibinfo{person}{Mahdieh Ahmadi}, \bibinfo{person}{Sokol Kosta}, {and}
  \bibinfo{person}{Pan Hui}.} \bibinfo{year}{2017}\natexlab{}.
\newblock \showarticletitle{Flopcoin: A cryptocurrency for computation
  offloading}.
\newblock \bibinfo{journal}{\emph{IEEE Transactions on Mobile Computing}}
  \bibinfo{volume}{17}, \bibinfo{number}{5} (\bibinfo{year}{2017}),
  \bibinfo{pages}{1062--1075}.
\newblock


\bibitem[\protect\citeauthoryear{Deng}{Deng}{2012}]%
        {deng2012mnist}
\bibfield{author}{\bibinfo{person}{Li Deng}.} \bibinfo{year}{2012}\natexlab{}.
\newblock \showarticletitle{The mnist database of handwritten digit images for
  machine learning research [best of the web]}.
\newblock \bibinfo{journal}{\emph{IEEE Signal Processing Magazine}}
  \bibinfo{volume}{29}, \bibinfo{number}{6} (\bibinfo{year}{2012}),
  \bibinfo{pages}{141--142}.
\newblock


\bibitem[\protect\citeauthoryear{He, Zhang, Ren, and Sun}{He
  et~al\mbox{.}}{2016}]%
        {he2016deep}
\bibfield{author}{\bibinfo{person}{Kaiming He}, \bibinfo{person}{Xiangyu
  Zhang}, \bibinfo{person}{Shaoqing Ren}, {and} \bibinfo{person}{Jian Sun}.}
  \bibinfo{year}{2016}\natexlab{}.
\newblock \showarticletitle{Deep residual learning for image recognition}. In
  \bibinfo{booktitle}{\emph{Proceedings of the IEEE conference on computer
  vision and pattern recognition}}. \bibinfo{pages}{770--778}.
\newblock


\bibitem[\protect\citeauthoryear{Hu, Jiang, and Wang}{Hu et~al\mbox{.}}{2019}]%
        {DBLP:journals/corr/abs-1908-07782}
\bibfield{author}{\bibinfo{person}{Chenghao Hu}, \bibinfo{person}{Jingyan
  Jiang}, {and} \bibinfo{person}{Zhi Wang}.} \bibinfo{year}{2019}\natexlab{}.
\newblock \showarticletitle{Decentralized Federated Learning: {A} Segmented
  Gossip Approach}.
\newblock \bibinfo{journal}{\emph{CoRR}}  \bibinfo{volume}{abs/1908.07782}
  (\bibinfo{year}{2019}).
\newblock
\showeprint[arxiv]{1908.07782}
\urldef\tempurl%
\url{http://arxiv.org/abs/1908.07782}
\showURL{%
\tempurl}


\bibitem[\protect\citeauthoryear{Koloskova, Stich, and Jaggi}{Koloskova
  et~al\mbox{.}}{2019}]%
        {koloskova2019decentralized}
\bibfield{author}{\bibinfo{person}{Anastasia Koloskova},
  \bibinfo{person}{Sebastian~U Stich}, {and} \bibinfo{person}{Martin Jaggi}.}
  \bibinfo{year}{2019}\natexlab{}.
\newblock \showarticletitle{Decentralized stochastic optimization and gossip
  algorithms with compressed communication}.
\newblock \bibinfo{journal}{\emph{arXiv preprint arXiv:1902.00340}}
  (\bibinfo{year}{2019}).
\newblock


\bibitem[\protect\citeauthoryear{{Latif}, {Khalifa}, {Rana}, and
  {Jurdak}}{{Latif} et~al\mbox{.}}{2020}]%
        {9111050}
\bibfield{author}{\bibinfo{person}{S. {Latif}}, \bibinfo{person}{S. {Khalifa}},
  \bibinfo{person}{R. {Rana}}, {and} \bibinfo{person}{R. {Jurdak}}.}
  \bibinfo{year}{2020}\natexlab{}.
\newblock \showarticletitle{Poster Abstract: Federated Learning for Speech
  Emotion Recognition Applications}. In \bibinfo{booktitle}{\emph{2020 19th
  ACM/IEEE International Conference on Information Processing in Sensor
  Networks (IPSN)}}. \bibinfo{pages}{341--342}.
\newblock


\bibitem[\protect\citeauthoryear{Ramanan, Nakayama, and Sharma}{Ramanan
  et~al\mbox{.}}{2019}]%
        {DBLP:journals/corr/abs-1909-07452}
\bibfield{author}{\bibinfo{person}{Paritosh Ramanan}, \bibinfo{person}{Kiyoshi
  Nakayama}, {and} \bibinfo{person}{Ratnesh Sharma}.}
  \bibinfo{year}{2019}\natexlab{}.
\newblock \showarticletitle{{BAFFLE} : Blockchain based Aggregator Free
  Federated Learning}.
\newblock \bibinfo{journal}{\emph{CoRR}}  \bibinfo{volume}{abs/1909.07452}
  (\bibinfo{year}{2019}).
\newblock
\showeprint[arxiv]{1909.07452}


\bibitem[\protect\citeauthoryear{Recht, Re, Wright, and Niu}{Recht
  et~al\mbox{.}}{2011}]%
        {recht2011hogwild}
\bibfield{author}{\bibinfo{person}{Benjamin Recht},
  \bibinfo{person}{Christopher Re}, \bibinfo{person}{Stephen Wright}, {and}
  \bibinfo{person}{Feng Niu}.} \bibinfo{year}{2011}\natexlab{}.
\newblock \showarticletitle{Hogwild: A lock-free approach to parallelizing
  stochastic gradient descent}. In \bibinfo{booktitle}{\emph{Advances in neural
  information processing systems}}. \bibinfo{pages}{693--701}.
\newblock


\bibitem[\protect\citeauthoryear{Robbins and Monro}{Robbins and Monro}{1951}]%
        {robbins1951stochastic}
\bibfield{author}{\bibinfo{person}{Herbert Robbins} {and}
  \bibinfo{person}{Sutton Monro}.} \bibinfo{year}{1951}\natexlab{}.
\newblock \showarticletitle{A stochastic approximation method}.
\newblock \bibinfo{journal}{\emph{The annals of mathematical statistics}}
  (\bibinfo{year}{1951}), \bibinfo{pages}{400--407}.
\newblock


\bibitem[\protect\citeauthoryear{Roy, Siddiqui, P{\"{o}}lsterl, Navab, and
  Wachinger}{Roy et~al\mbox{.}}{2019}]%
        {DBLP:journals/corr/abs-1905-06731}
\bibfield{author}{\bibinfo{person}{Abhijit~Guha Roy}, \bibinfo{person}{Shayan
  Siddiqui}, \bibinfo{person}{Sebastian P{\"{o}}lsterl},
  \bibinfo{person}{Nassir Navab}, {and} \bibinfo{person}{Christian Wachinger}.}
  \bibinfo{year}{2019}\natexlab{}.
\newblock \showarticletitle{BrainTorrent: {A} Peer-to-Peer Environment for
  Decentralized Federated Learning}.
\newblock \bibinfo{journal}{\emph{CoRR}}  \bibinfo{volume}{abs/1905.06731}
  (\bibinfo{year}{2019}).
\newblock
\showeprint[arxiv]{1905.06731}


\bibitem[\protect\citeauthoryear{Yang, Liu, Chen, and Tong}{Yang
  et~al\mbox{.}}{2019}]%
        {10.1145/3298981}
\bibfield{author}{\bibinfo{person}{Qiang Yang}, \bibinfo{person}{Yang Liu},
  \bibinfo{person}{Tianjian Chen}, {and} \bibinfo{person}{Yongxin Tong}.}
  \bibinfo{year}{2019}\natexlab{}.
\newblock \showarticletitle{Federated Machine Learning: Concept and
  Applications}.
\newblock \bibinfo{journal}{\emph{ACM Trans. Intell. Syst. Technol.}}
  \bibinfo{volume}{10}, \bibinfo{number}{2}, Article \bibinfo{articleno}{12}
  (\bibinfo{date}{Jan.} \bibinfo{year}{2019}), \bibinfo{numpages}{19}~pages.
\newblock
\showISSN{2157-6904}


\bibitem[\protect\citeauthoryear{Yang, Andrew, Eichner, Sun, Li, Kong, Ramage,
  and Beaufays}{Yang et~al\mbox{.}}{2018}]%
        {DBLP:journals/corr/abs-1812-02903}
\bibfield{author}{\bibinfo{person}{Timothy Yang}, \bibinfo{person}{Galen
  Andrew}, \bibinfo{person}{Hubert Eichner}, \bibinfo{person}{Haicheng Sun},
  \bibinfo{person}{Wei Li}, \bibinfo{person}{Nicholas Kong},
  \bibinfo{person}{Daniel Ramage}, {and} \bibinfo{person}{Fran{\c{c}}oise
  Beaufays}.} \bibinfo{year}{2018}\natexlab{}.
\newblock \showarticletitle{Applied Federated Learning: Improving Google
  Keyboard Query Suggestions}.
\newblock \bibinfo{journal}{\emph{CoRR}}  \bibinfo{volume}{abs/1812.02903}
  (\bibinfo{year}{2018}).
\newblock
\showeprint[arxiv]{1812.02903}


\end{thebibliography}
\end{document}